\documentclass{revtex4}
\usepackage{slashed}
\usepackage{epsfig, subfigure}
\usepackage{graphicx}
\usepackage{fancyhdr}
\usepackage{amsmath}
\def\ga{\mathrel{\raise.3ex\hbox{$>$\kern-.75em\lower1ex\hbox{$\sim$}}}}
\def\la{\mathrel{\raise.3ex\hbox{$<$\kern-.75em\lower1ex\hbox{$\sim$}}}}
\usepackage{color}

\begin{document}

\title{2HDM at the LHC - the story so far}

\author{A. Barroso}
\affiliation{Centro de F\'{\i}sica Te\'{o}rica e Computacional,
    Faculdade de Ci\^{e}ncias,
    Universidade de Lisboa,
    Av.\ Prof.\ Gama Pinto 2,
    1649-003 Lisboa, Portugal}
\author{P.M.~Ferreira}
    \email[E-mail: ]{ferreira@cii.fc.ul.pt}
\affiliation{Instituto Superior de Engenharia de Lisboa - ISEL,
	1959-007 Lisboa, Portugal}
\affiliation{Centro de F\'{\i}sica Te\'{o}rica e Computacional,
    Faculdade de Ci\^{e}ncias,
    Universidade de Lisboa,
    Av.\ Prof.\ Gama Pinto 2,
    1649-003 Lisboa, Portugal}
\author{Rui Santos}
    \email[E-mail: ]{rsantos@cii.fc.ul.pt}
\affiliation{Instituto Superior de Engenharia de Lisboa - ISEL,
	1959-007 Lisboa, Portugal}
\affiliation{Centro de F\'{\i}sica Te\'{o}rica e Computacional,
    Faculdade de Ci\^{e}ncias,
    Universidade de Lisboa,
    Av.\ Prof.\ Gama Pinto 2,
    1649-003 Lisboa, Portugal}
\author{Marc Sher}
\affiliation{ High Energy Theory Group, College of William and Mary,  Williamsburg, Virginia 23187, U.S.A.}
\author{Jo\~{a}o P.~Silva}
    \email[E-mail: ]{jpsilva@cftp.ist.utl.pt}
\affiliation{Instituto Superior de Engenharia de Lisboa,
	1959-007 Lisboa, Portugal}
\affiliation{Centro de F\'{\i}sica Te\'{o}rica de Part\'{\i}culas (CFTP),
    Instituto Superior T\'{e}cnico, Universidade T\'{e}cnica de Lisboa,
    1049-001 Lisboa, Portugal}

\begin{abstract}
We confront the most common version of the CP-conserving 2HDM with LHC data, taking into account all previously 
available experimental data. We also discuss the scenario where the 125 GeV Higgs discovered at the LHC is the lightest 
neutral scalar of a particular CP-violating 2HDM. In this scenario we focus on what data can already tell us about 
the amount of mixing between CP-even and CP-odd states.
\end{abstract}

\maketitle

\thispagestyle{fancy}

\section{Introduction}

After the discovery~\cite{ATLASHiggs, CMSHiggs} of a Higgs boson at CERN's Large Hadron Collider (LHC), it is now time to
confront possible extensions of the Standard Model (SM) with LHC data. One of the 
simplest ways to extend the scalar sector of the SM is to add one more complex doublet to the model.
The resulting two Higgs doublet models (2HDMs) can provide additional CP-violation
coming from the scalar sector and can easily originate dark matter candidates. More evolved
models with additional field content have a 2HDM like scalar sector - as in the case, for instance,
of the Minimal Supersymmetric Standard Model. 
2HDMs have a richer particle spectrum with one charged and three neutral
scalars. All neutral scalars could in principle be the scalar discovered at the 
LHC~\cite{Ferreira:2011aa, Ferreira:2012my, Burdman:2011ki}. However, data  
gathered so far allow us to rule out the 125 GeV particle as being a pure pseudoscalar state. Hence,
we are left with either one of the CP-even states of a CP-conserving model or a neutral 
state from a CP-violating model~\cite{Barroso:2012wz}. In the next section we will present the 
2HDM models
and in the following sections we discuss in turn how well  2HDMs can
accommodate   LHC data for the CP-conserving and for the CP-violating case.

\section{The Models}
\label{sec:models}

In this section we present the two versions of the 2HDM to be discussed, one CP-conserving
and the other explicitly CP-violating. The most general 2HDMs' Yukawa Lagrangian
gives rise to scalar exchange flavour changing neutral currents (FCNCs)
which are strongly constrained by experiment. A simple and natural way to 
avoid those dangerous FCNCs is to impose a $Z_2$ symmetry on the
scalar doublets, $\Phi_1 \rightarrow \Phi_1$,
$\Phi_2 \rightarrow - \Phi_2$,  and a corresponding symmetry to the
quark fields. This leads to the well known four Yukawa model types I, II, Y (III, Flipped)
and X (IV, Lepton Specific)~\cite{barger, KY}. The different Yukawa types are built such that 
only $\phi_2$ couples to all fermions (type I), or $\phi_2$ couples to up-type quarks and $\phi_1$ couples to 
down-type quarks and leptons (type II), or $\phi_2$ couples to up-type quarks and 
to leptons and $\phi_1$ couples to down-type quarks (type Y) or finally $\phi_2$ couples to all quarks and $\phi_1$ couples to leptons (type X).

The scalar potential in a softly broken $Z_2$ symmetric 2HDM can be written as
\begin{align*}
V(\Phi_1,\Phi_2) =& m^2_1 \Phi^{\dagger}_1\Phi_1+m^2_2
\Phi^{\dagger}_2\Phi_2 + (m^2_{12} \Phi^{\dagger}_1\Phi_2+{\mathrm{h.c.}
}) +\frac{1}{2} \lambda_1 (\Phi^{\dagger}_1\Phi_1)^2 +\frac{1}{2}
\lambda_2 (\Phi^{\dagger}_2\Phi_2)^2\nonumber \\ 
+& \lambda_3
(\Phi^{\dagger}_1\Phi_1)(\Phi^{\dagger}_2\Phi_2) + \lambda_4
(\Phi^{\dagger}_1\Phi_2)(\Phi^{\dagger}_2\Phi_1) + \frac{1}{2}
\lambda_5[(\Phi^{\dagger}_1\Phi_2)^2+{\mathrm{h.c.}}] ~, \label{higgspot}
\end{align*}
where $\Phi_i$, $i=1,2$ are complex SU(2) doublets. All parameters except for 
$m_{12}^2$ and $\lambda_5$ are real as a consequence of the hermiticity
of the potential. The phases of $m_{12}^2$ and $\lambda_5$ together with 
the ones in the vacuum expectation values (VEVs) will determine the CP nature
of the model (see~\cite{Branco:2011iw} for a review). A CP-violating model
has three spin 0 neutral states usually denoted by $h_1$,  $h_2$ and $h_3$
while in the CP-conserving case the three CP-eigenstates are usually denoted
by $h$ and $H$ (CP-even) and $A$ (CP-odd).
As shown in~\cite{vacstab1, vacstab2}, once either a CP-conserving or a CP-violating 
vacuum configuration is chosen, all charge breaking stationary points are saddle
points with higher energy.  Hence, the 2HDM is stable at tree-level against charge 
breaking and once a non-charge breaking vacuum is chosen both models 
have two charged Higgs bosons that
complete the 2HDM particle spectrum.

We will focus on two specific realisations of 2HDMs. One is the usual 7-parameter 
CP-conserving potential where $m_{12}^2$, $\lambda_5$ and the VEVs are all real.
In this model we choose as free parameters, the four masses, the rotation
angle in the CP-even sector, $\alpha$, the ratio of the vacuum expectation
values,  $\tan\beta=v_2/v_1$, and the soft breaking parameter redefined as $M^2=m_{12}^2/(\sin \beta \, \cos \beta)$.
The other is an explicit CP-violating model~\cite{Ginzburg:2002wt,ElKaffas:2006nt, Arhrib:2010ju}. 
In the CP-violating version $m_{12}^2$ and $\lambda_5$ are complex and 
the fields' VEVs are real. Existence of a stationary point
requires $Im (m_{12}^2) = v_1 \, v_2 \, Im ( \lambda_5)$. Because the VEVs
are real in both models, a common definition for the rotation angle
in the charged sector $\tan\beta=v_2/v_1$ can be used. 
In this model, besides $m_{H^\pm}$ and $\tan \beta$, we take as free parameters
the masses of the lightest scalars $m_{h1}$ and $m_{h2}$ the three rotation
angles in the neutral sector $\alpha_1$, $\alpha_2$ and $\alpha_3$
and $Re(m_{12}^2)$.
Details of the
model can be found in~\cite{Ginzburg:2002wt,ElKaffas:2006nt, Arhrib:2010ju}.

\subsection{Constraints on the models}

We have imposed the following theoretical bounds on both models: we require that the potential is bounded
from below~\cite{vac1} and we impose unitarity limits on the quartic Higgs couplings~\cite{unitarity}. 
We have also taken into account the precision electroweak constraints~\cite{Peskin:1991sw, STHiggs}.
Furthermore, we discuss the scenarios in which two normal minima coexist in the CP-conserving potential. In
those scenarios it could happen that we are living in a local minimum and tunnelling to a 
global deeper minimum could occur. This deeper minimum would have a different $v^2$ and therefore
different masses for the elementary particles. We call this a panic vacuum~\cite{vacstab2, IvanonPRE, Barroso:2012mj}. 
The experimental bounds pertaining to the charged sector apply to both models since the charged
Higgs Yukawa couplings have exactly the same form. Except for the LEP bounds they all
constrain regions of the $(\tan \beta, m_{H^\pm}) $ plane. The LEP experiments have set a lower
limit on the mass of the charged Higgs boson of 80 GeV  at 95\% C.L., assuming
$BR(H^+ \to \tau^+ \nu) + BR(H^+ \to c \bar s) +  BR(H^+ \to A W^+) =1$~\cite{LEP2013}
with the process $e^+ e^- \to H^+ H^-$. The bound
on the mass is 94 GeV if $BR(H^+ \to \tau^+ \nu)  =1$~\cite{LEP2013}.  
These are model independent bounds as long as the above mentioned sum
of BRs is one. Values of $\tan \beta$ smaller than $O (1)$ together with 
a charged Higgs with a mass below $O (100$ GeV$)$ are both disallowed by the constraints~\cite{BB} coming from
$R_b$, from $B_q \bar{B_q}$ 
mixing and  from $B\to X_s \gamma$~\cite{BB2} for all models. Furthermore, data from $B\to X_s \gamma$~\cite{BB2}
 imposes a lower limit of $m_{H^\pm}  \ga 360$ GeV, but only for models type II and type Y.
We have also considered the most recent bounds from the ATLAS~\cite{ATLASICHEP} and CMS~\cite{CMSICHEP} collaborations
on the $(\tan \beta, m_{H^\pm}) $ plane coming from $pp \to t \bar{t} \to b \bar{b} W^\mp H^\pm (\to \tau \nu)$.
Finally, we have considered the LEP bounds on the neutral Higgs~\cite{mssmhiggs}.

Combining it all,
our scans will be performed taking  $m_{H^\pm} \geq 90$ GeV and $\tan \beta \geq 1$ for type I
while  $m_{H^\pm} > 360$ GeV for type II. For the CP-conserving case we also take $m_A \geq 90$ GeV, $m_h = 125$ GeV and
$m_H > m_h$ while $\alpha$ is free to vary in its allowed range although subject to the above constraints.
For the CP-violating model we take $m_{h1} = 125$ GeV and
$m_{h2} > m_{h1}$ while $\alpha_i$ ($i=1,2,3$) are again free to vary in their allowed
ranges, subject to the above constraints. 
Finally,
we note that we do not include the constraint that would result
from the anomaly observed by the BaBar collaboration
in the rates $R (D)$ and $R(D^*)$,
with $R(D) = (\overline{B}\to D \tau^-\overline{\nu}_\tau)/(\overline{B}\to D l^-\overline{\nu}_l)$,
which deviates by 3.4~$\sigma$ (when $D$ and $D^*$ final states are
combined) from the SM prediction~\cite{Lees:2012xj}.
It is clear that the deviation cannot be explained by a charged Higgs boson
from either of the four FCNC conserving Yukawa types. Therefore, if this observation is independently confirmed
by the BELLE collaboration not only the SM will be excluded at 3.4~$\sigma$ but all
2HDMs Yukawa types will also be excluded with a very similar significance.

We should point out that there will never be an exclusion of the
2HDM in favor of the SM.
Indeed,
the models discussed here include the SM as a specific (decoupling)
region of parameter space.
As a result,
experiments may exclude the SM in favor of the 2HDM;
they may exclude both;
or they may be compatible with SM,
thus restricting the 2HDM extension into a particular
(eventually, small) region of the parameter space.


\section{CP-conserving model}

We start with the CP-conserving model and randomly generate points in the
parameter space such that $m_h = 125$ GeV, 90 GeV   $ \leq m_A \leq$ 900 GeV,
$m_h < m_H \leq$ 900 GeV, $1 \leq \tan \beta \leq 40$, $- (900)^2$ GeV$^2$ 
$ \leq m_{12}^2 \leq 900^2$ GeV$^2$ and $-\pi/2 \leq \alpha \leq \pi/2$.
Further, for type I we take 90 GeV $ \leq m_{H^\pm} \leq$ 900 GeV while
for type II the allowed range is 360 GeV $ \leq m_{H^\pm} \leq$ 900 GeV. 
In order to use the LHC data we define the quantities $R_f$ as the ratio of the number of events
predicted in the 2HDM to that obtained in the SM for a given final state $f$.
\begin{equation} \label{Sratio}
R_f =
\frac{\sigma(pp \to h)_{\textrm{2HDM}}\
\textrm{BR}(h \to f)_{\textrm{2HDM}}}{
\sigma(pp \to h_{\textrm{SM}})\ \textrm{BR}(h_{\textrm{SM}} \to f)},
\end{equation}
where $h$ is the lightest CP-even Higgs (125 GeV), $\sigma$ is the Higgs production
cross section, BR the branching ratio,
and $h_{\textrm{SM}}$ is the SM Higgs boson.
In our analysis,
we include all Higgs production mechanisms,
namely,
gluon-gluon fusion using HIGLU at NLO~\cite{Spira:1995mt},
vector boson fusion (VBF)~\cite{LHCHiggs}, Higgs production in
association with either $W$, $Z$  or
$t\bar{t}$~\cite{LHCHiggs},
and  $b \bar{b}$ fusion~\cite{Harlander:2003ai}. A different approach
was recently followed in~\cite{Chiang:2013ixa} where they have looked for the best fit point 
in parameter space for the four Yukawa types
of the same CP-conserving 2HDM.

\begin{figure}[h!]
\centering
\includegraphics[width=3.3in,angle=0]{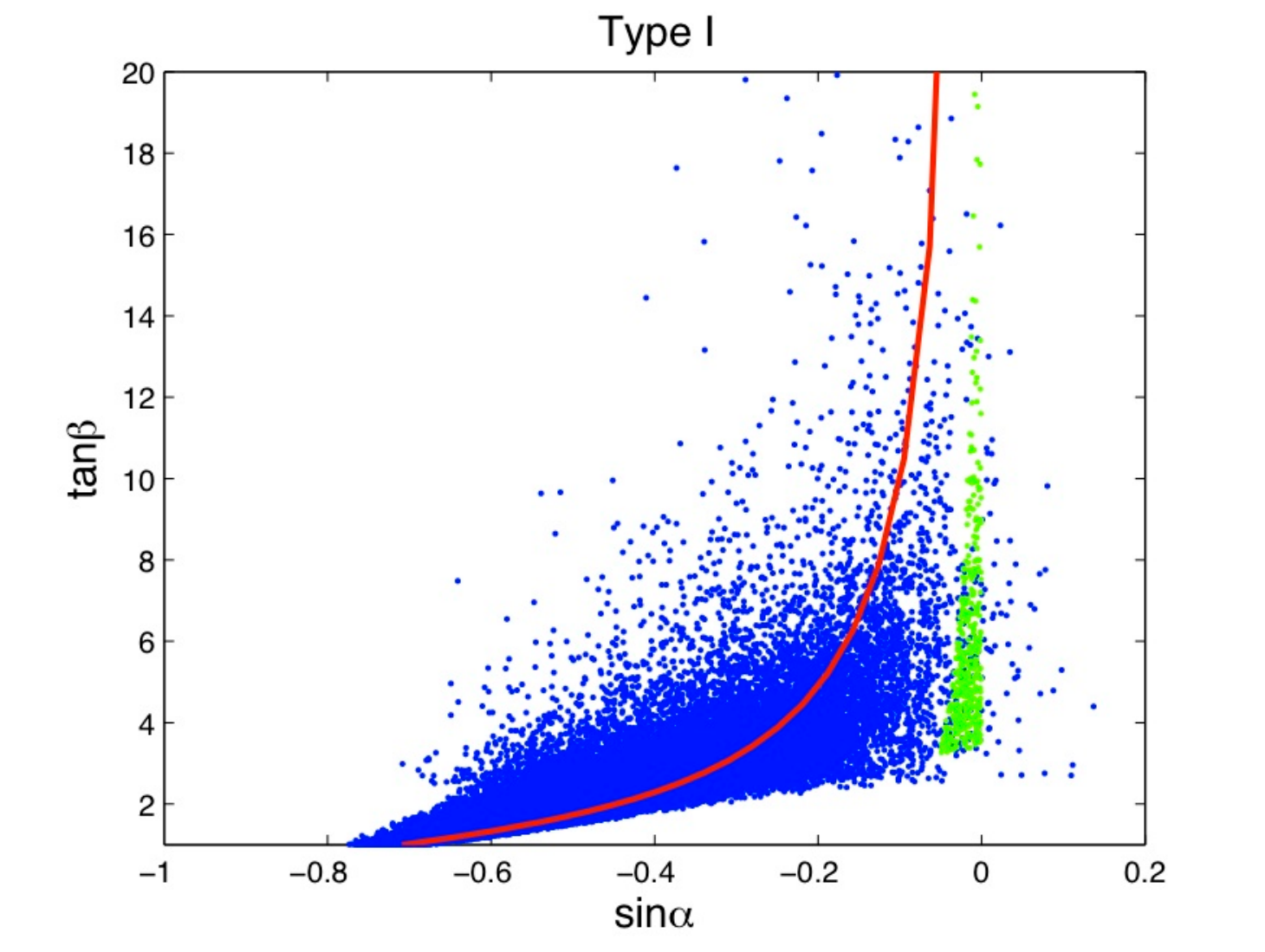}
\includegraphics[width=3.3in,angle=0]{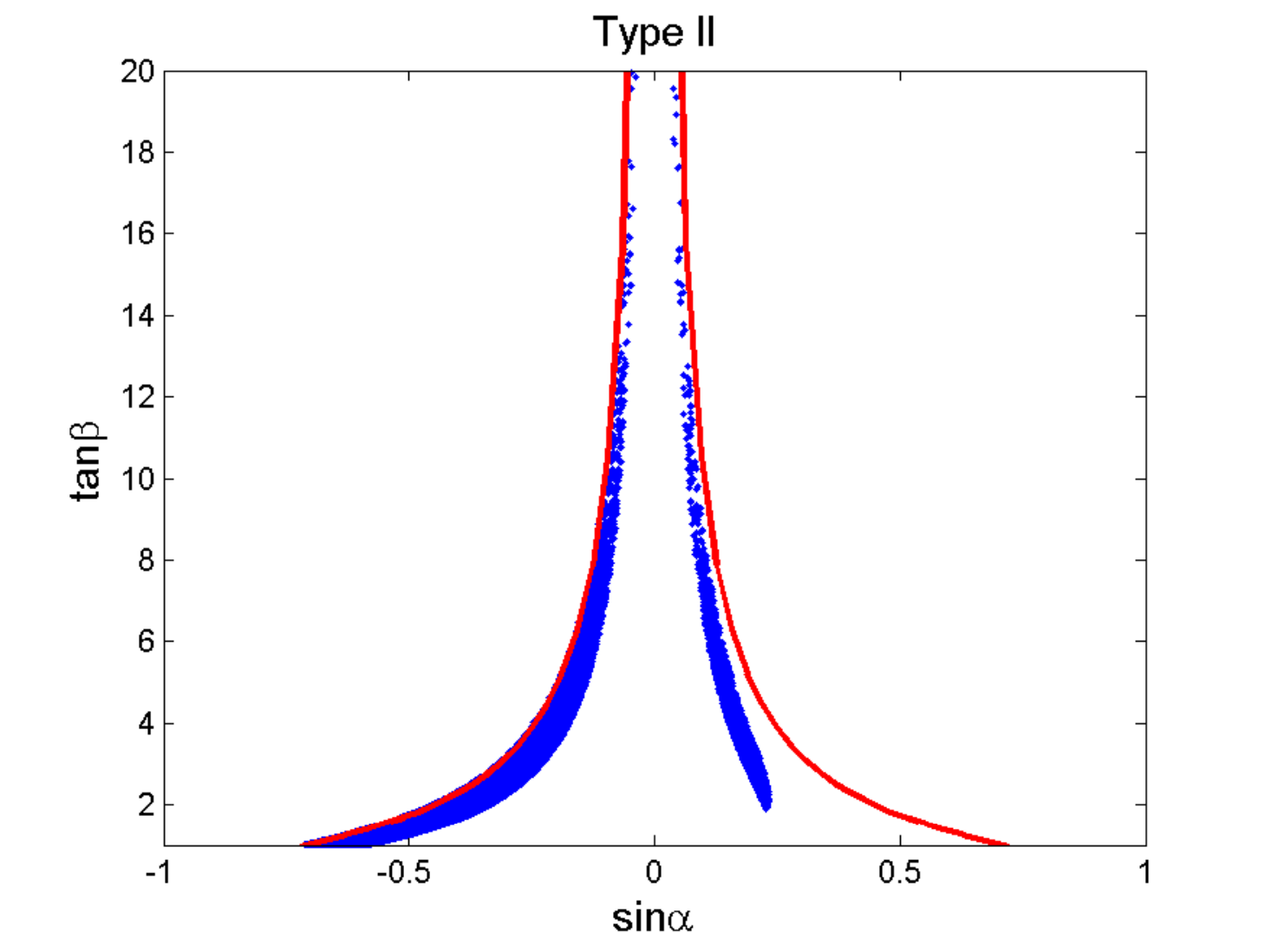}
\vspace{-0.3cm}
\caption{Points in the  ($\sin \alpha$, $\tan \beta$)  plane that passed all
 the constraints in type I (left) and in type II (right) at 1$\sigma$ in green (light grey) and 2$\sigma$ in blue (dark grey). 
Also shown are the
lines for the SM-like limit, that is $\sin(\beta - \alpha) =1$ (negative $\sin \alpha$) and for the limit
$\sin(\beta + \alpha) =1$ (positive $\sin \alpha$).}
\label{fig1}
\end{figure}

The points
generated have then to pass all the constraints previously described plus
the combined~\cite{Arbey:2012bp} ATLAS and CMS strengths $R_{\gamma \gamma} = 1.66 \pm 0.33$,
$R_{ZZ} = 0.93 \pm 0.28$ and  $R_{ \tau \tau} = 0.71 \pm 0.42$.    It is important to point out that this talk was given in February, 2013, prior to the updates from Moriond.   We have added a subsection regarding the Moriond updates below.
In figure~\ref{fig1} we present the points in the   ($\sin \alpha$, $\tan \beta$)  plane that pass
all the constraints in type I (left) and in type II (right) at 1$\sigma$ in green and at 2$\sigma$ in blue. 
Also shown are the lines for the SM-like limit, that is $\sin(\beta - \alpha) =1$ (negative $\sin \alpha$) and for the limit
$\sin(\beta + \alpha) =1$ (positive $\sin \alpha$). 

To better understand these results we approximate
$R_{ZZ}$ in the case where Higgs production is due exclusively to gluon-gluon fusion
via the top quark loop,
and the total Higgs width is well approximated by $\Gamma(h \to b \bar{b})$.
This is a very good approximation for most of the parameter space and it yields 
\begin{equation}
R^{I, approx}_{ZZ} =
\frac{\cos^2{\alpha}}{\sin^2{\beta}}
\sin^2{(\alpha - \beta)} \frac{\sin^2{\beta}}{\cos^2{\alpha}}
=
\sin^2{(\alpha - \beta)}.
\label{RZZ_I_simplified}
\end{equation}
By setting $R^{I, approx}_{ZZ} = 1$ one obtains the SM-like limit line
shown in the picture for negative $\alpha$.
Applying the same simplified scenario we obtain for type II
\begin{equation}
R^{II, approx}_{ZZ} =
\sin^2{(\alpha - \beta)} \frac{1}{\tan^2{\alpha} \, \tan^2{\beta}}.
\label{RZZ_II_simplified}
\end{equation}
%
Now, when $R^{II, approx}_{ZZ} = 1$ we obtain not one but two lines, the red lines
shown in figure~\ref{fig1}. The  approximate scenarios lead us to the following 
conclusions. First, it is very hard in type I to have $R_{ZZ}$ above 1. Further,
the function $\sin^2{(\alpha - \beta)}$ is very sensitive to deviations from 1. This leads
to a larger dispersion of points around the SM-like limit line. Regarding type II, not only do
we get two lines instead of one but the function  $R^{II, approx}_{ZZ}$ is now very
insensitive to deviations from 1 and that is why the points are all in a narrow band
close to each line. It is interesting to note that while the tension between $R_{ZZ}$
and $R_{\gamma \gamma}$ excludes all points at 1$\sigma$ in type II, there is still
a band in type I close to $\sin \alpha =0$ where $R_{ZZ}$ is below 1 while $R_{\gamma \gamma}$
can be large. The 1$\sigma$ points in type I all have a charged Higgs mass below 130 GeV.

\begin{figure}[h!]
\centering
\includegraphics[width=3.2in,angle=0]{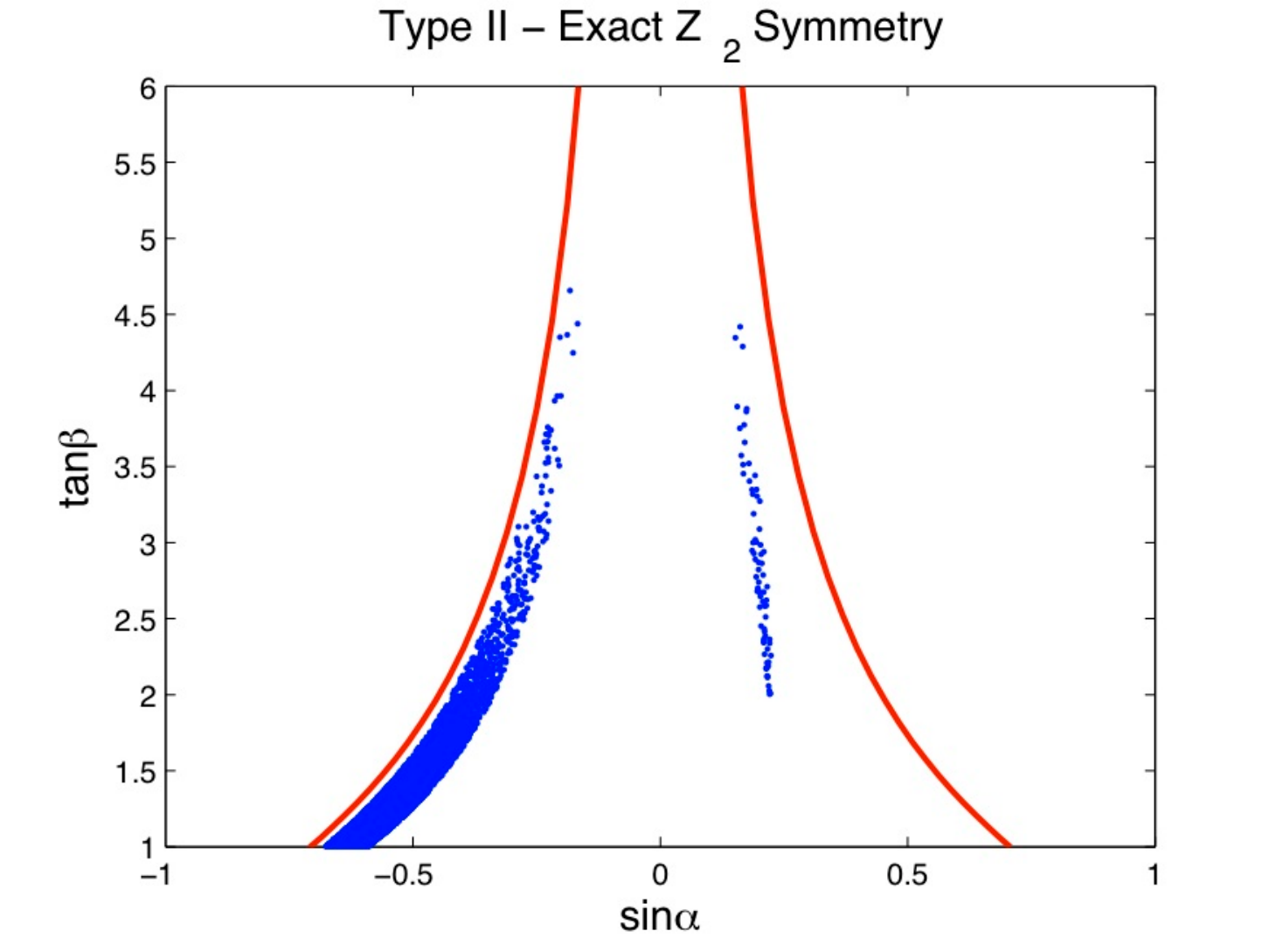}
\includegraphics[width=3.2in,angle=0]{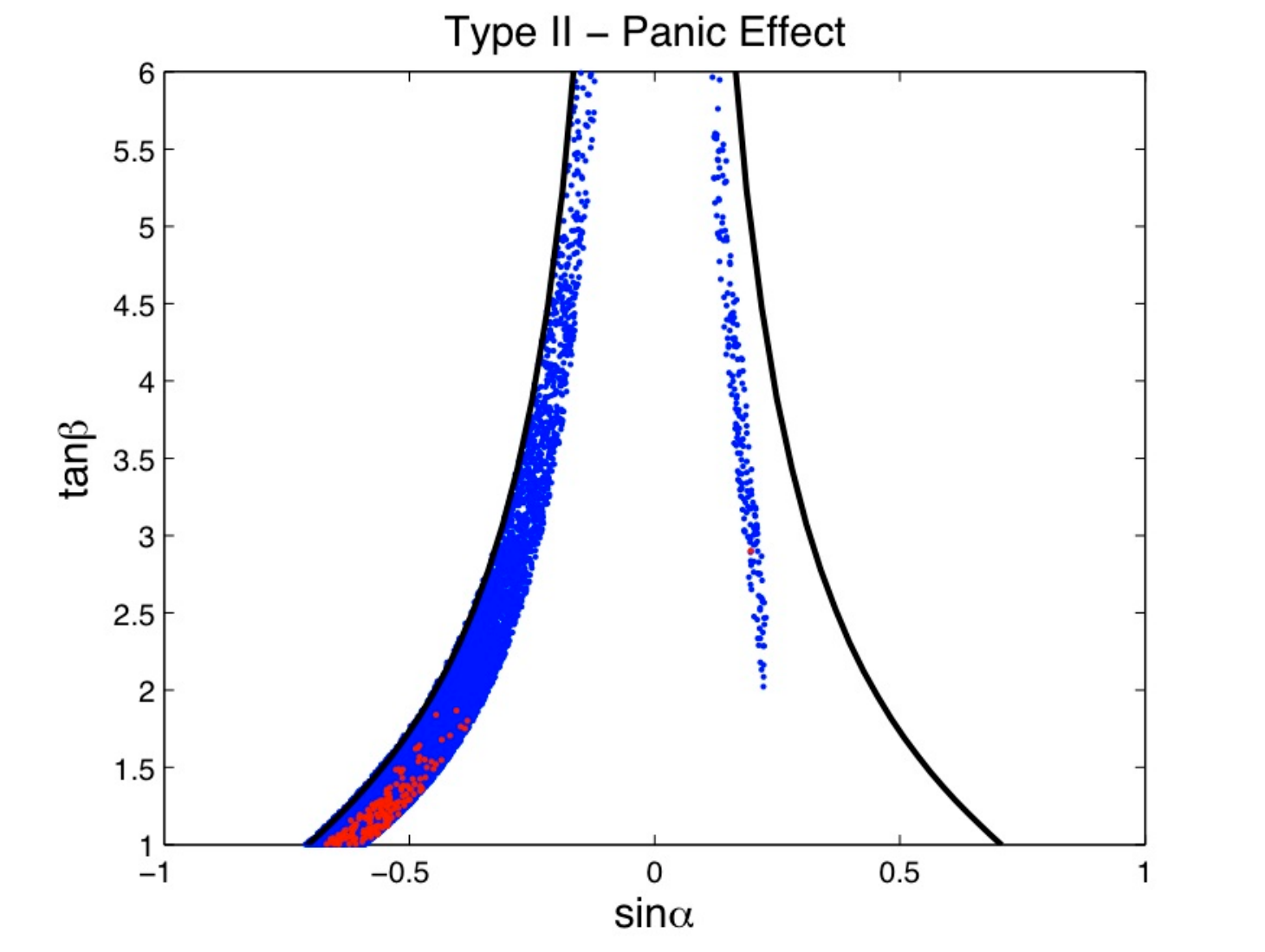}
\vspace{-0.3cm}
\caption{In the left panel we show the points in the  ($\sin \alpha$, $\tan \beta$) plane that have passed all
 the constraints in type II with an exact $Z_2$ symmetry. In the right panel we show type II with the broken $Z_2$
 symmetry where we have plotted the panic points in red. 
No points survive at 1$\sigma$ and the 2$\sigma$ points are shown in blue.}
\label{fig2}
\end{figure}

\begin{figure}[h!]
\centering
\includegraphics[width=3.3in,angle=0]{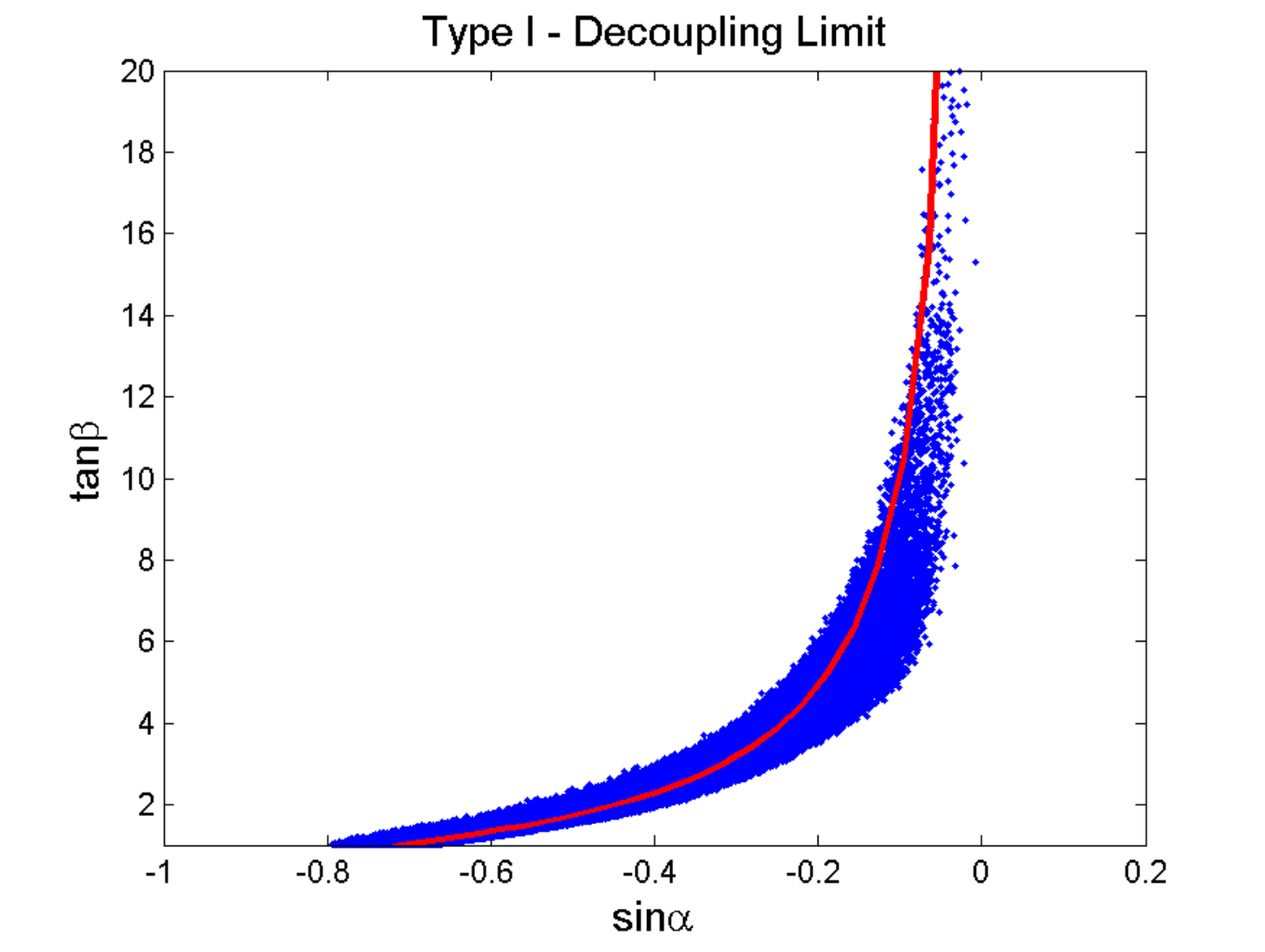}
\includegraphics[width=3.3in,angle=0]{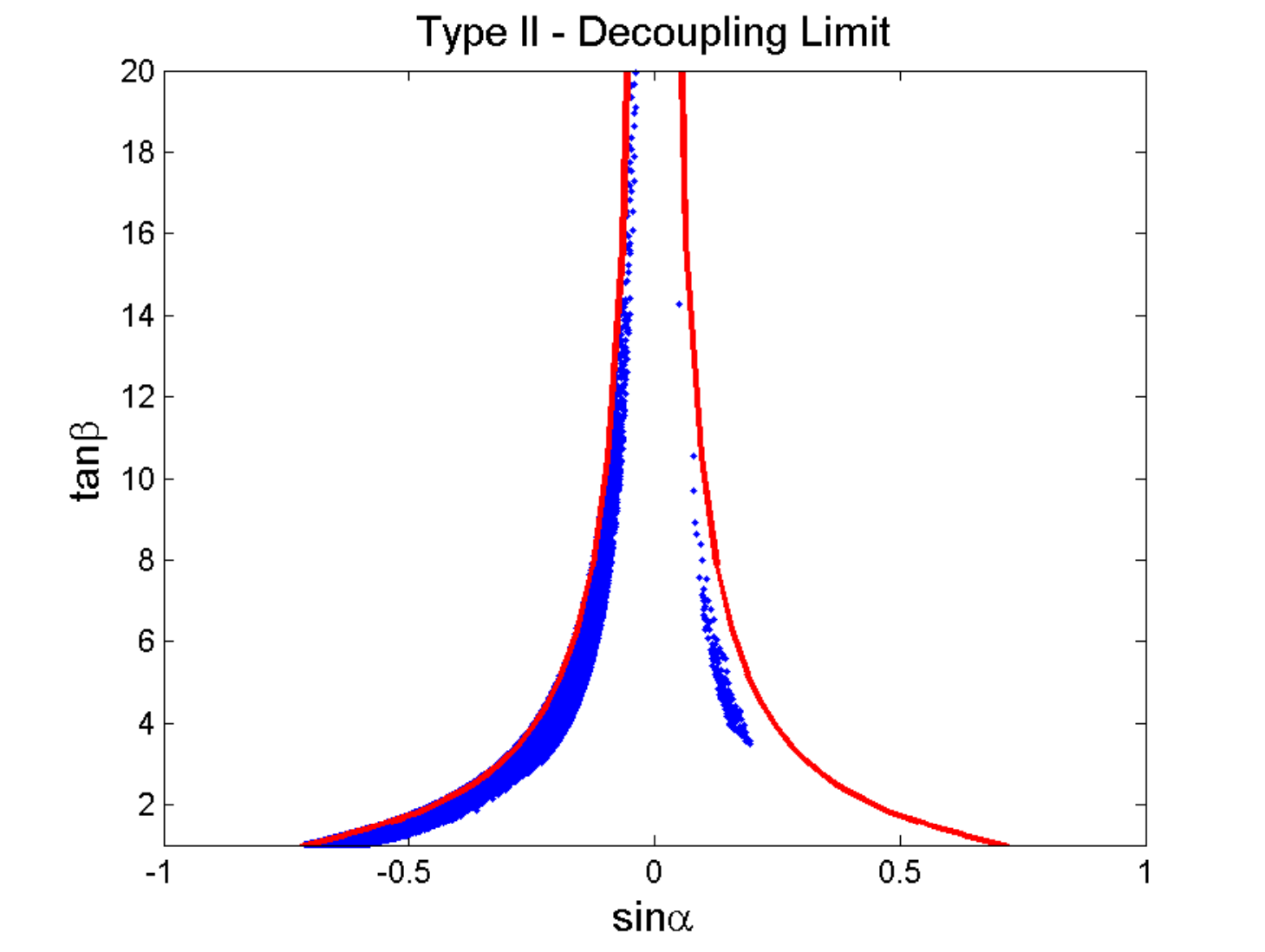}
\vspace{-0.3cm}
\caption{Points in the  ($\sin \alpha$, $\tan \beta$) plane that passed all
 the constraints in type I (left) and in type II (right) at 2$\sigma$ in blue. We have
 taken all masses except $m_h =125$ GeV to be above 600 GeV.}
\label{fig3}
\end{figure}

In the left panel of figure~\ref{fig2} we present the points that have passed all the constraints in type II
with an exact $Z_2$ symmetry ($m_{12}^2 =0$). It was recently shown~\cite{Gorczyca:2011he} that if $m_h =125$ GeV, then $\tan \beta$
has to be below approximately 6 .  This is confirmed in the plot. Apart from the limit on $\tan \beta$, the plot
does not differ much from the softly broken $Z_2$ case except that
there are fewer points and that they move slightly 
away from the red lines when compared with the soft breaking scenario. 
We note that in type I no points survive at 1$\sigma$ in the 
exact $Z_2$ symmetric case.
In the right panel of  figure~\ref{fig2}
we show the panic points (in red) for the soft breaking type II model.  Panic points are easily found by
using the following discriminant~\cite{Barroso:2012mj}
\begin{equation}
D =
\left( m_{11}^2 - k^2 m_{22}^2 \right)
(\tan{\beta} - k).
\label{D}
\end{equation}
where $k=\sqrt[4]{\lambda_1/\lambda_2}$. If $D<0$ we live in a metastable state and  our current $(v_1, v_2)$
solution is the panic vacuum.
We found panic points distributed all over
the 2$\sigma$ allowed region. We should note however that as shown in~\cite{Barroso:2012mj}, the panic points are further away
from the SM-like limit than the non-panic points. 

In figure~\ref{fig3} we present similar plots to the ones shown in figure~\ref{fig1} the only difference being that
all masses except for the lightest Higgs one are taken to be above 600 GeV. This plot is representative of the decoupling limit
of the 2HDM - if all particles except $h$ are too heavy to be detected at the LHC this is how the plot would look 
after the bounds on those masses are considered.

We finish this section with a comment on the values of $M^2$ and the values of the remaining masses that are still allowed by data.
We had already noted that the points with $M^2=0$ are disallowed at 2$\sigma$ in type I. We have checked that 
also $M^2<0$ is excluded at 2$\sigma$ in type I. In type II all values of $M^2$ are allowed at 2$\sigma$.
Regarding the remaining masses,
 our conclusion is that the data on the 125 GeV Higgs does not provide meaningful
constraints on the masses of the remaining scalars.

\subsection{Update after Moriond 2013}

In this section we present updated plots after Moriond 2013. We use the latest updates from
from ATLAS~\cite{ATLASnotes} and from CMS~\cite{CMSnotes}  for the 7 and 8 TeV runs.
Regarding the CMS results for $h \to \gamma \gamma$ we use the result from the Multivariate
analysis.

\begin{figure}[h!]
\centering
\includegraphics[width=3.2in,angle=0]{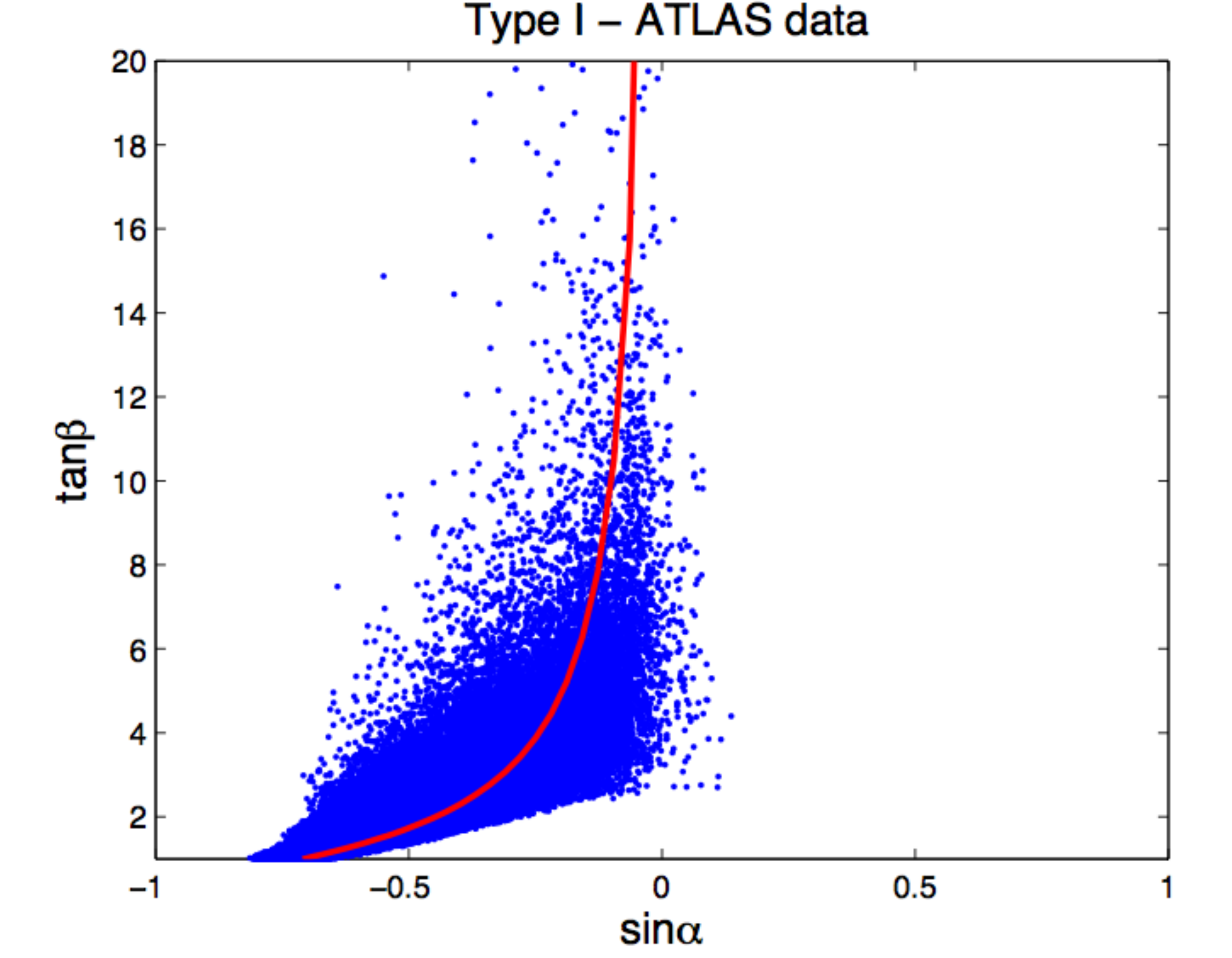}
\includegraphics[width=3.4in,angle=0]{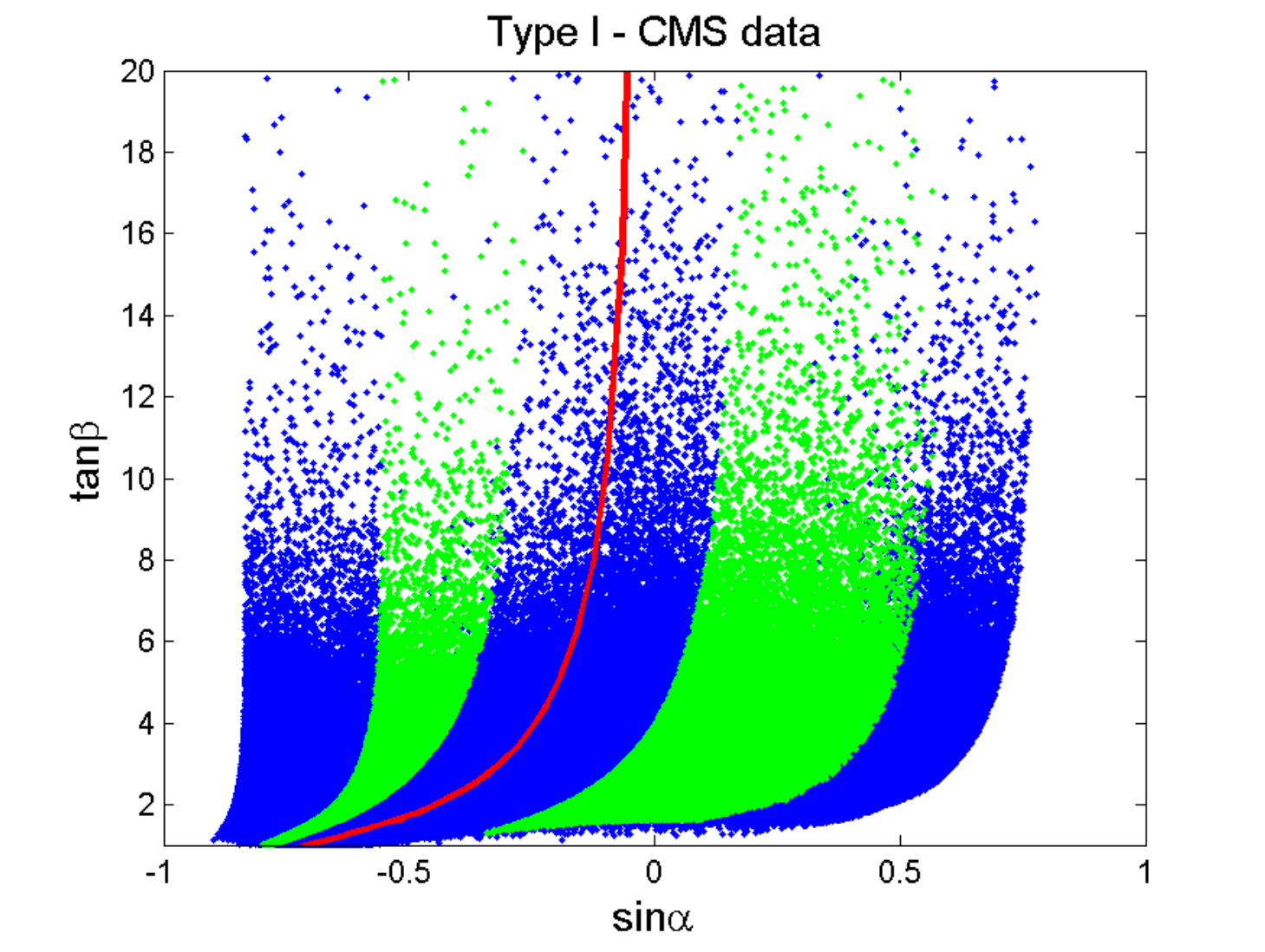}
\vspace{-0.3cm}
\caption{Points in the  ($\sin \alpha$, $\tan \beta$)  plane that passed all
 the constraints in type I using the ATLAS data (left) and using the CMS data (right) at 1$\sigma$ in green (light grey) and 2$\sigma$ in blue (dark grey). 
Also shown is the
line for the SM-like limit, $\sin(\beta - \alpha) =1$.}
\label{fig4}
\end{figure}

\begin{figure}[h!]
\centering
\includegraphics[width=3.3in,angle=0]{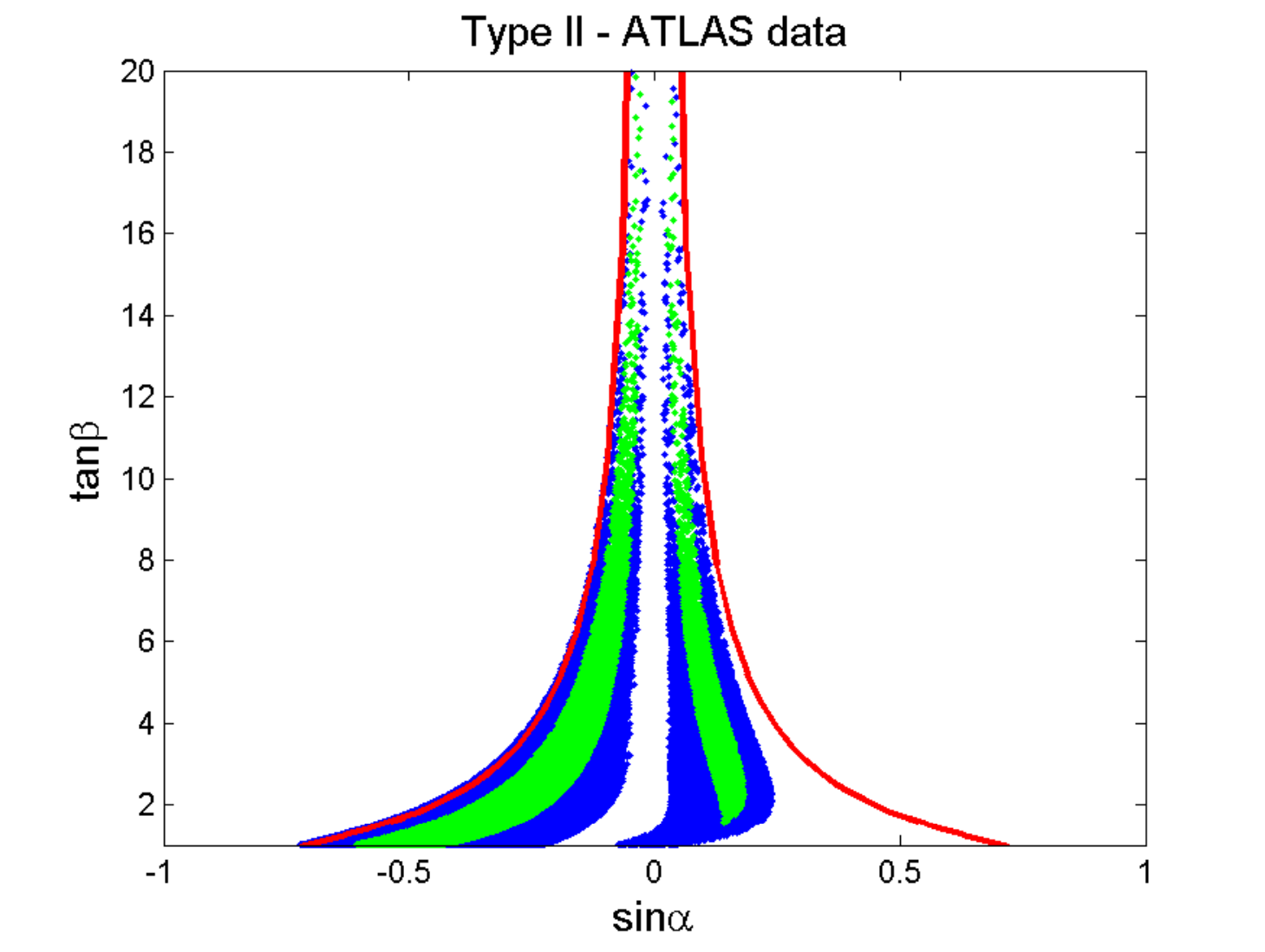}
\includegraphics[width=3.3in,angle=0]{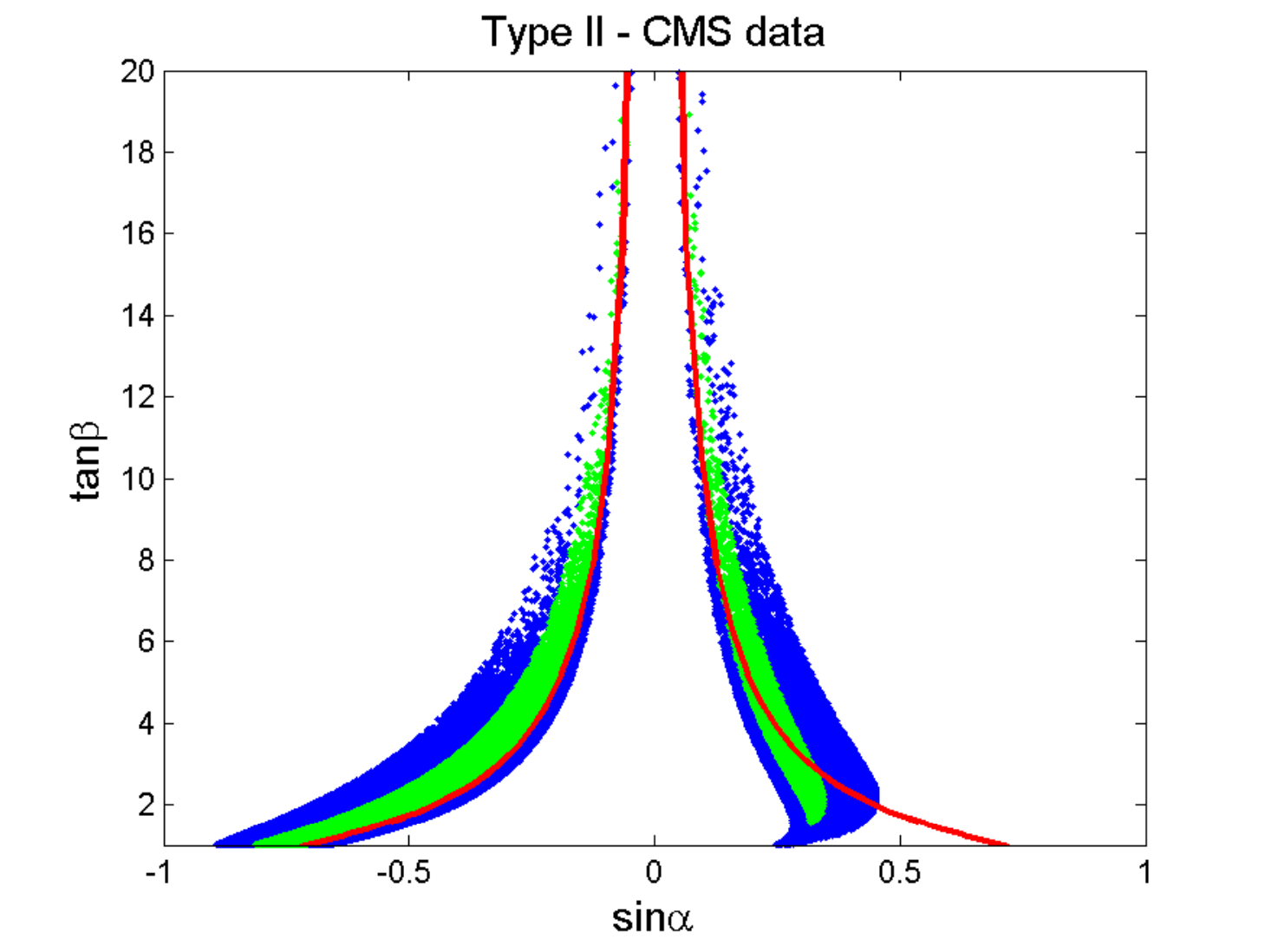}
\vspace{-0.3cm}
\caption{Points in the  ($\sin \alpha$, $\tan \beta$)  plane that passed all
 the constraints in type II using the ATLAS data (left) and using the CMS data (right) at 1$\sigma$ in green (light grey) and 2$\sigma$ in blue (dark grey). 
 Also shown are the
lines for the SM-like limit, that is $\sin(\beta - \alpha) =1$ (negative $\sin \alpha$) and for the limit
$\sin(\beta + \alpha) =1$ (positive $\sin \alpha$).}
\label{fig5}
\end{figure}

In figure~\ref{fig4} we present the results for the type I model  in the ($\sin \alpha$, $\tan \beta$)  plane
using the ATLAS data (left) and the CMS data (right).
The differences between the two plots are easy to understand. The ATLAS data forces $R_{ZZ}$ to be large
but as we have previously shown (equation~\ref{RZZ_I_simplified})  $R_{ZZ}$ can never be above one in type I. Consequently,
no points survive at 1$\sigma$. With the CMS data plenty of 1$\sigma$ points survive because all $R_{VV}$ are below one.
As expected,
the 1$\sigma$ region is slightly away from the SM-like limit because,
on one hand,
the central values of $R_{VV}$ are below one,
and,
on the other hand,
we already saw that $R_{VV}$ are very sensitive quantities in type I.

\begin{figure}[h!]
\centering
\includegraphics[width=3.3in,angle=0]{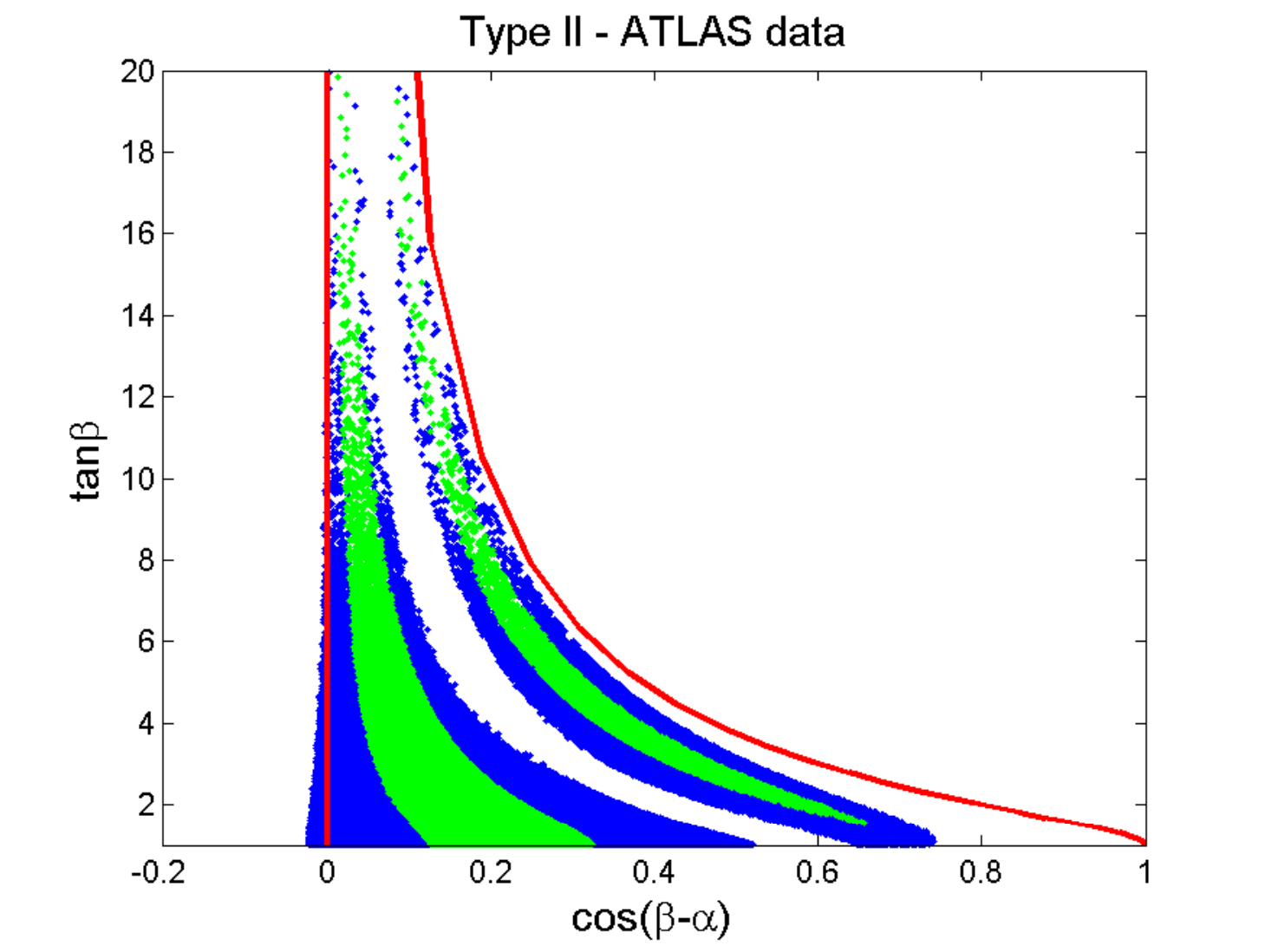}
\includegraphics[width=3.3in,angle=0]{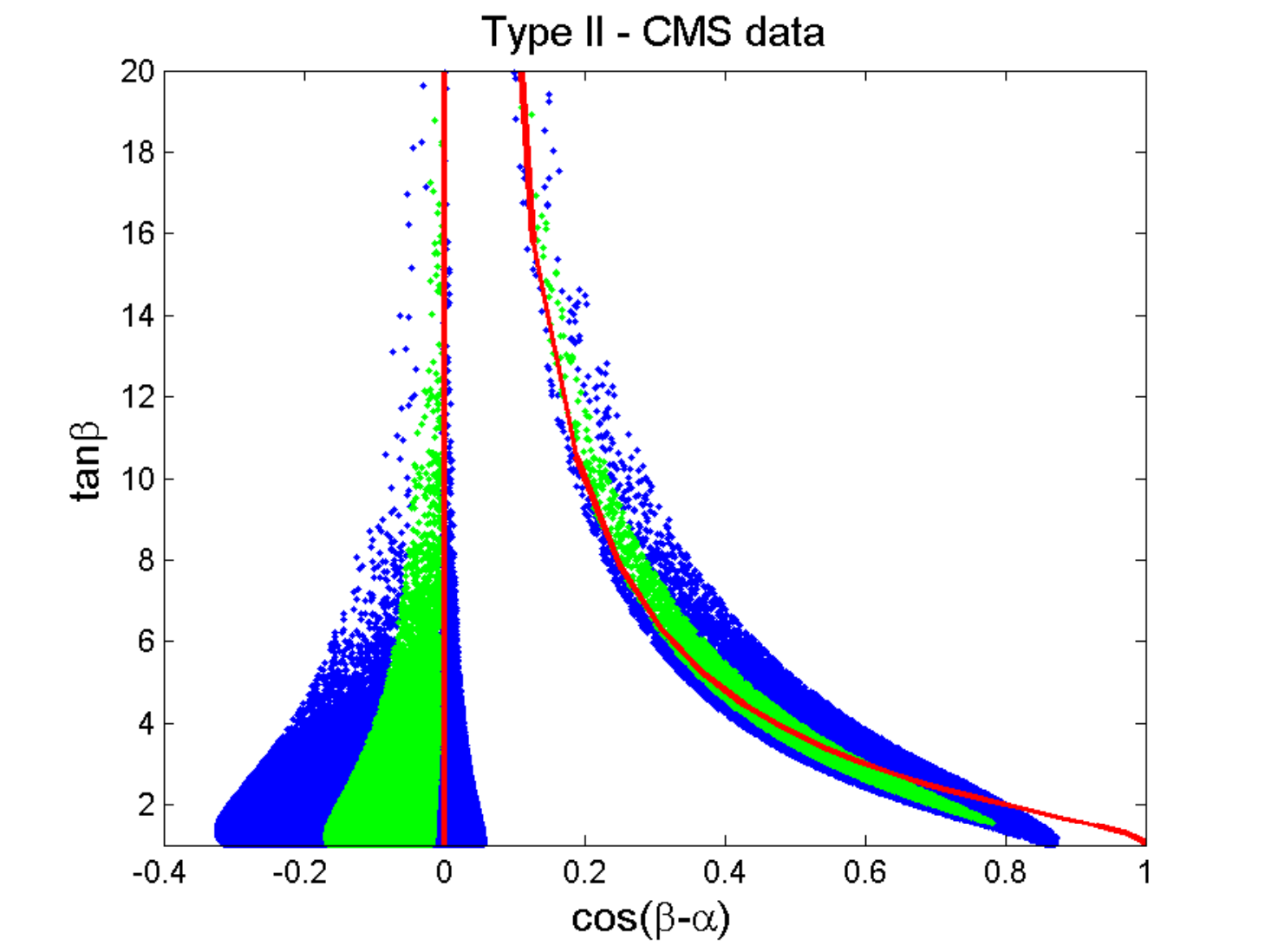}
\vspace{-0.3cm}
\caption{Points in the  ($\cos (\beta - \alpha)$, $\tan \beta$)  plane that passed all
 the constraints in type II using the ATLAS data (left) and using the CMS data (right) at 1$\sigma$ in green (light grey) and 2$\sigma$ in blue (dark grey). 
 Also shown are the
lines for the SM-like limit, that is $\cos (\beta - \alpha) =0$ and for the limit
$\cos(\beta + \alpha) =0$.}
\label{fig6}
\end{figure}

In figure~\ref{fig5} we now show similar plots for the type II model. In type II $R_{VV}$ is not a sensitive quantity. Therefore
both the 1$\sigma$ and the 2$\sigma$ points are very close to the two limiting lines (in red). The difference between the two plots
is that the points tend to concentrate below the limiting lines for ATLAS and above the same lines for CMS. This is of course
a consequence of the central values of the ATLAS $R_{VV}$ being above the SM while the CMS ones are below the SM expectation. 
Finally we present in figure~\ref{fig6} plots similar to the ones in figure~\ref{fig5} but now in the ($\cos (\beta - \alpha)$, $\tan \beta$)
plane. Again we see the same trend with the allowed points on opposite sides of the limiting lines, $\cos (\beta - \alpha) =0$ and
$\cos(\beta + \alpha) =0$.

\section{CP-violating model}

In this section we discuss the CP-violating model 2HDM presented in section~\ref{sec:models}.
The lightest neutral particle of the model $h_1$ is considered to be the one discovered
at the LHC. This particle can be seen as being composed by a CP-even and a CP-odd 
component. The amount of mixing is controlled
by $s_2 = \sin{\alpha_2}$
(one of the three rotation angles in the neutral sector - see~\cite{Barroso:2012wz} for details). 
When $s_2 = 0$, $h_1$ is a pure CP-even state and when $s_2 = 1$, $h_1$ is a pure CP-odd state. Our goal
is to constrain the amount of mixing using LHC data.
\begin{figure}[h!]
\centering
\includegraphics[width=3.3in,angle=0]{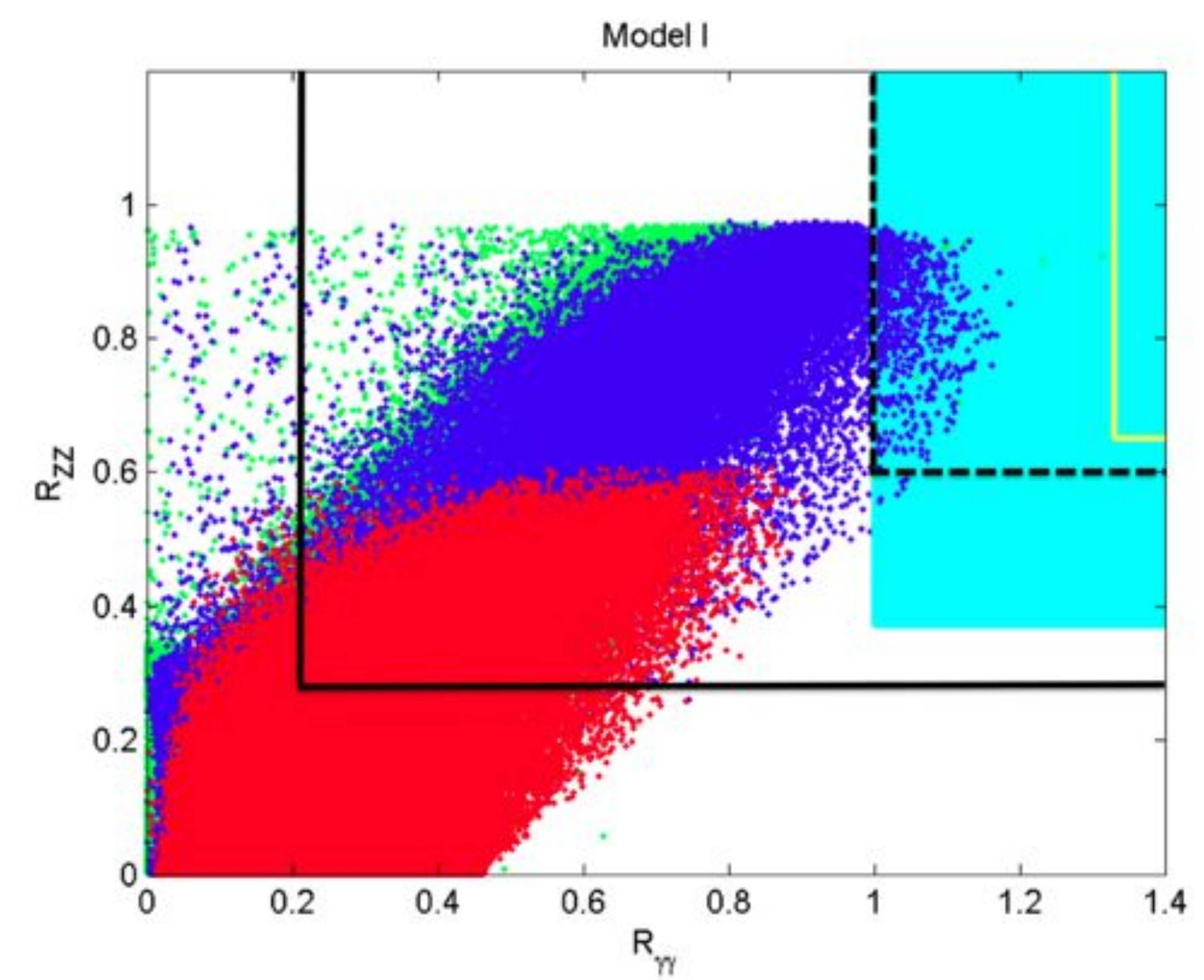}
\includegraphics[width=3.3in,angle=0]{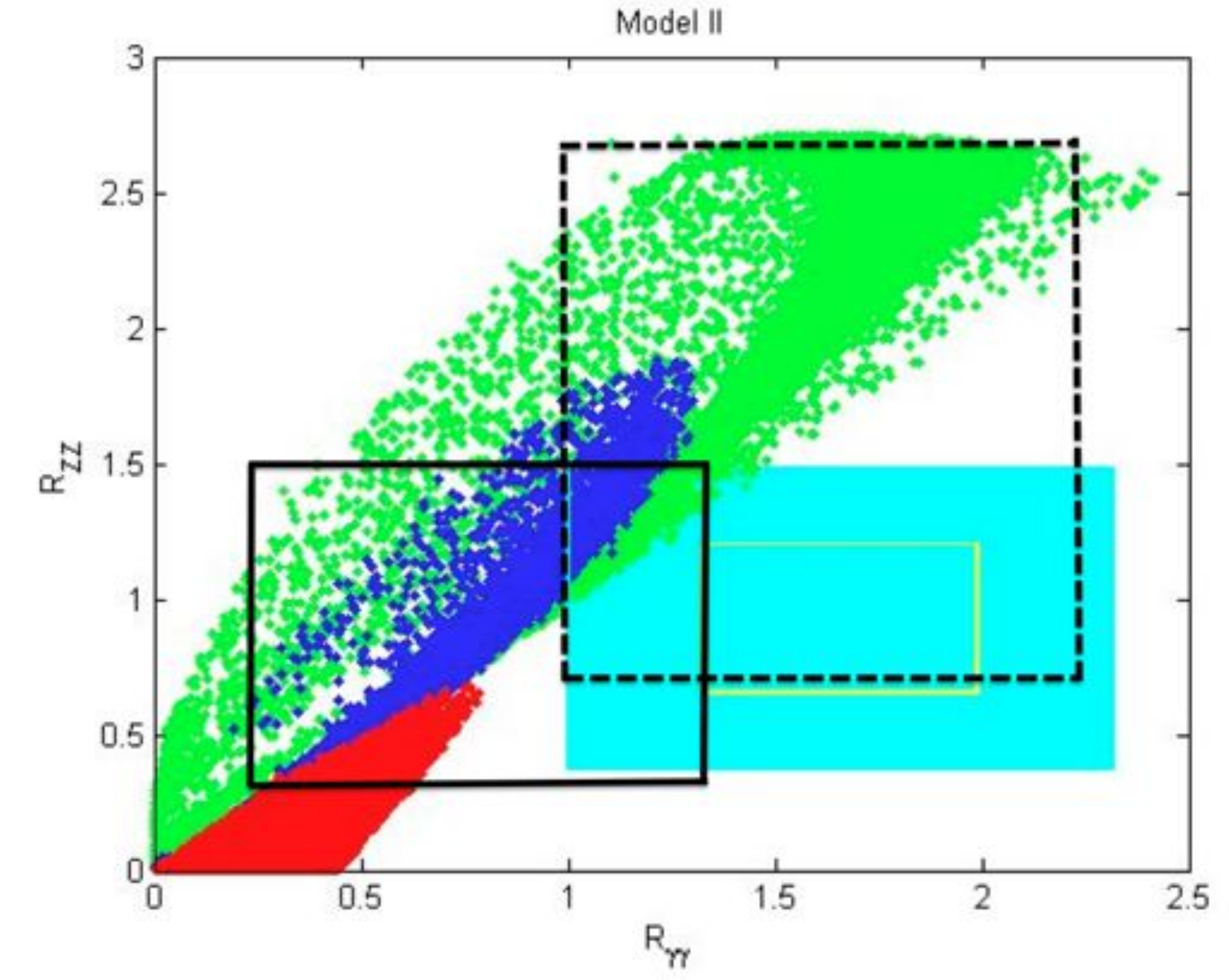}
\vspace{-0.3cm}
\caption{In the left panel we show $R_{ZZ}$ versus $R_{\gamma \gamma}$ for type I
and for three regions of the parameter $s_2$: $|s_2| < 0.1$ in green
(light grey), $0.45 < |s_2| < 0.55$ in blue (black)
and $|s_2| > 0.83$ in red (dark grey). In the right panel we present a similar plot
for type II.
We show in both plots the combined ATLAS and CMS results at 1$\sigma$ (yellow line) and at 2$\sigma$ in blue and the
2$\sigma$ updated lines after Moriond 2013 (solid black - CMS and dashed black - ATLAS). }
\label{fig7}
\end{figure}
In the left panel of figure~\ref{fig7} we show $R_{ZZ}$ versus $R_{\gamma \gamma}$ for type I
and for three regions of the parameter $s_2$: $|s_2| < 0.1$ in green (light grey),
$0.45 < |s_2| < 0.55$ in blue (black),
and $|s_2| > 0.83$ in red (dark grey).
In the right panel we present a similar plot
for type II.
We also show in both plots the combined ATLAS and CMS results at 1$\sigma$ (yellow line) and at 2$\sigma$ in blue.
We conclude that the red points are excluded 
at 2$\sigma$ when the combined results~\cite{Arbey:2012bp} from ATLAS and CMS are considered.

The results of the Moriond updates are shown as 2$\sigma$
solid (CMS) and dashed (ATLAS) lines in figure~\ref{fig7}.
We see that $|s_2| > 0.83$ is still excluded at 2$\sigma$
by the ATLAS data, but that it is allowed by the new CMS data.

\section{Conclusions}

We have considered the most common CP-conserving 2HDMs and a CP-violating 2HDM in light of the recent data from the LHC.   For the CP conserving case, 
there are 4 different models, corresponding to the transformation properties of the quarks, leptons and Higgs bosons under a $Z_2$ symmetry.  
 We include constraints from vacuum stability, unitarity, precision electroweak constraints, LEP experimental bounds, bounds from b decays as 
 well as LHC bounds on top decays.   The full parameter space is scanned, subject to these constraints, and allowed points are plotted in 
 the $(\sin\alpha,\tan\beta)$ plane and in the $(\sin\alpha,\cos (\beta - \alpha))$ plane.   This talk was presented before the Moriond 2013 meeting, at which time the Standard Model 
 branching ratios disagreed by more than one sigma with the combined data from ATLAS and CMS, and thus the plots show no points within  one sigma of the 
 Standard Model curves.   However, we have updated the plots since Moriond, and find that in type I CMS data does allow points within one 
 sigma of the Standard Model.  In type II both ATLAS and CMS now allow points within one sigma of the SM.
  In the next run, these experimental bounds will tighten considerably, providing a strong test of the models.

In the CP violating case, the key parameter is $s_2$, which controls the amount by which the observed $125$ GeV state is CP-even vs. CP-odd.    
We have shown that some regions of relatively large $s_2$ can be excluded, but a wide range of values remains.   
Over the next few years, the rectangles shown in figure~\ref{fig7}  will shrink considerably, providing a much stronger bound 
on the amount of the scalar's CP-odd component.

\begin{acknowledgments}
RS is grateful to the workshop organisation for financial support and for providing the 
opportunity for very stimulating discussions.
The work of AB, PMF and RS is supported in part by the Portuguese
\textit{Funda\c{c}\~{a}o para a Ci\^{e}ncia e a Tecnologia} (FCT)
under contract PTDC/FIS/117951/2010, by FP7 Reintegration Grant, number PERG08-GA-2010-277025,
and by PEst-OE/FIS/UI0618/2011.
The work of JPS is funded by FCT through the projects
CERN/FP/109305/2009 and  U777-Plurianual,
and by the EU RTN project Marie Curie: PITN-GA-2009-237920.  The work of MS is supported by the NSF under Grant No.~PHY-1068008.

\end{acknowledgments}


\begin{thebibliography}{99} 

\bibitem{ATLASHiggs}
G.~Aad {\it et al.}  [ATLAS Collaboration],
  Phys.\ Lett.\ B {\bf 716}, 1 (2012)
  [arXiv:1207.7214 [hep-ex]].

\bibitem{CMSHiggs}
S.~Chatrchyan {\it et al.}  [CMS Collaboration],
  Phys.\ Lett.\ B {\bf 716}, 30 (2012)
  [arXiv:1207.7235 [hep-ex]].

\bibitem{Ferreira:2011aa} 
  P.~M.~Ferreira, R.~Santos, M.~Sher, J.~P.~Silva,
  Phys.\ Rev.\ D {\bf 85}, 077703 (2012)
  [arXiv:1112.3277 [hep-ph]].
  
\bibitem{Ferreira:2012my} 
  P.~M.~Ferreira, R.~Santos, M.~Sher, J.~P.~Silva,
  Phys.\ Rev.\ D {\bf 85}, 035020 (2012)
  [arXiv:1201.0019 [hep-ph]].
  
\bibitem{Burdman:2011ki} 
  G.~Burdman, C.~E.~F.~Haluch, R.~D.~Matheus,
  Phys.\ Rev.\ D {\bf 85}, 095016 (2012)
  [arXiv:1112.3961 [hep-ph]].
  
\bibitem{Barroso:2012wz} 
  A.~Barroso, P.~M.~Ferreira, R.~Santos, J.~P.~Silva,
  Phys.\ Rev.\ D {\bf 86}, 015022 (2012)
  [arXiv:1205.4247 [hep-ph]].

\bibitem{barger}
V.~D.~Barger, J.~L.~Hewett and R.~J.~N.~Phillips,
  Phys. \ Rev.\  D {\bf 41} (1990) 3421.

\bibitem{KY}
  M.~Aoki, S.~Kanemura, K.~Tsumura and K.~Yagyu,
  Phys.\ Rev.\  D {\bf 80} (2009) 015017.

\bibitem{Branco:2011iw}
  G.~C.~Branco, P.~M.~Ferreira, L.~Lavoura, M.~N.~Rebelo, M.~Sher and J.~P.~Silva,
  Phys.\ Rept.\  {\bf 516} (2012) 1
  [arXiv:1106.0034 [hep-ph]].

\bibitem{vacstab1}
P.~M.~Ferreira, R.~Santos and A.~Barroso,
  Phys.\ Lett.\  B {\bf 603} (2004) 219
  [Erratum-ibid.\  B {\bf 629} (2005) 114];
  A.~Barroso, P.~M.~Ferreira, R.~Santos,
  Phys.\ Lett.\ B {\bf 632}, 684 (2006)
  [hep-ph/0507224];
 %
%
  M.~Maniatis, A.~von Manteuffel, O.~Nachtmann, F.~Nagel,
  Eur.\ Phys.\ J.\ C {\bf 48}, 805 (2006)
  [hep-ph/0605184].
\bibitem{vacstab2}
  I.~P.~Ivanov,
  Phys.\ Rev.\ D {\bf 75}, 035001 (2007)
  [Erratum-ibid.\ D {\bf 76}, 039902 (2007)]
  [hep-ph/0609018];
%
  I.~P.~Ivanov,
  Phys.\ Rev.\ D {\bf 77}, 015017 (2008)
  [arXiv:0710.3490 [hep-ph]].



\bibitem{Ginzburg:2002wt}
  I.~F.~Ginzburg, M.~Krawczyk and P.~Osland,
hep-ph/0211371.

\bibitem{ElKaffas:2006nt}
  A.~W.~El Kaffas, W.~Khater, O.~M.~Ogreid and P.~Osland,
  Nucl.\ Phys.\ B {\bf 775} (2007) 45
  [hep-ph/0605142].

%
\bibitem{Arhrib:2010ju}
  A.~Arhrib, E.~Christova, H.~Eberl and E.~Ginina,
JHEP {\bf 1104} (2011) 089.


\bibitem{vac1}
  N.G.~Deshpande and E.~Ma,
  Phys.\ Rev.\  D {\bf 18} (1978) 2574.


\bibitem{unitarity}
S.~Kanemura, T.~Kubota and E.~Takasugi,
Phys.\ Lett.\  B {\bf 313} (1993)  155; 
A.G.~Akeroyd, A.~Arhrib and E.M.~Naimi,
  Phys.\ Lett.\  B {\bf 490} (2000)  119.

\bibitem{Peskin:1991sw}
  M.E.~Peskin and T.~Takeuchi,
  Phys.\ Rev.\ D {\bf 46}, 381 (1992).

\bibitem{STHiggs}
 H.E.~Haber,
  ``Introductory Low-Energy Supersymmetry,'' in
\textit{Recent directions in particle theory: from superstrings and black holes to the standard model}, Proceedings of
the Theoretical Advanced Study Institute (TASI 92), Boulder, CO, 1--26
June 1992, edited by J.~Harvey and J.~Polchinski (World Scientific
Publishing, Singapore, 1993) pp.~589--688;
  C.~D.~Froggatt, R.~G.~Moorhouse and I.~G.~Knowles,
  Phys.\ Rev.\  D {\bf 45}, 2471 (1992);
 W.~Grimus, L.~Lavoura, O.~M.~Ogreid and P.~Osland,
  Nucl.\ Phys.\ B {\bf 801}, 81 (2008)
  [arXiv:0802.4353 [hep-ph]];
  H.E.~Haber and D.~O'Neil,
  Phys.\ Rev.\ D {\bf 83}, 055017 (2011)
  [arXiv:1011.6188 [hep-ph]].

\bibitem{IvanonPRE} 
I.~P.~Ivanov, 
 Phys.\ Rev.\ E {\bf 79}, 021116 (2009).

\bibitem{Barroso:2012mj} 
  A.~Barroso, P.~M.~Ferreira, I.~P.~Ivanov, R.~Santos, J.~P.~Silva,
  arXiv:1211.6119 [hep-ph];
%
  A.~Barroso, P.~M.~Ferreira, I.~P.~Ivanov, R.~Santos,
  arXiv:1303.5098 [hep-ph].

\bibitem{LEP2013} 
  G.~Abbiendi {\it et al.}  [ALEPH and DELPHI and L3 and OPAL and The LEP working group for Higgs boson searches Collaborations],
  [arXiv:1301.6065 [hep-ex]].


\bibitem{BB}

  A.~Denner, R.J.~Guth, W.~Hollik and J.H.~Kuhn,
  Z.\ Phys.\ C {\bf 51}, 695 (1991).
%
%
  H.E.~Haber and H.E.~Logan,
  Phys.\ Rev.\ D {\bf 62}, 015011 (2000)
  [hep-ph/9909335].
%
The ALEPH, CDF,  D0, DELPHI, L3, OPAL, SLD Collaborations, the LEP Electroweak Working Group, the Tevatron Electroweak Working Group, and the SLD electroweak and heavy flavour Groups,
  arXiv:1012.2367 [hep-ex].
  F.~Mahmoudi and O.~Stal,
  Phys.\ Rev.\ D {\bf 81}, 035016 (2010)
  [arXiv:0907.1791 [hep-ph]];
%
  M.~Baak, M.~Goebel, J.~Haller, A.~Hoecker, D.~Ludwig, K.~Moenig, M.~Schott and J.~Stelzer,
  Eur.\ Phys.\ J.\ C {\bf 72}, 2003 (2012)
  [arXiv:1107.0975 [hep-ph]].
%
M.~Baak, M.~Goebel, J.~Haller, A.~Hoecker, D.~Kennedy, R.~Kogler, K.~Moenig, M.~Schott and J.~Stelzer,
  arXiv:1209.2716 [hep-ph].
  A.~Wahab El Kaffas, P.~Osland, O.~M.~Ogreid,
  Phys.\ Rev.\ D {\bf 76}, 095001 (2007)
  [arXiv:0706.2997 [hep-ph]];
  L.~Basso, A.~Lipniacka, F.~Mahmoudi, S.~Moretti, P.~Osland, G.~M.~Pruna, M.~Purmohammadi,
  JHEP {\bf 1211}, 011 (2012)
  [arXiv:1205.6569 [hep-ph]].


\bibitem{BB2}
  D.~Asner {\it et al.}  [Heavy Flavor Averaging Group Collaboration],
  arXiv:1010.1589 [hep-ex].
%
T.~Hermann, M.~Misiak and M.~Steinhauser,
  JHEP {\bf 1211} (2012) 036.
See also,
%
F. Mahmoudi, talk given at Prospects For Charged Higgs Discovery At Colliders
(CHARGED 2012), 8-11 October, Uppsala, Sweden.

\bibitem{ATLASICHEP}
  G.~Aad {\it et al.}  [ATLAS Collaboration],
  JHEP {\bf 1206} (2012) 039
  [arXiv:1204.2760 [hep-ex]];
\bibitem{CMSICHEP}
  S.~Chatrchyan {\it et al.}  [CMS Collaboration],
  JHEP {\bf 1207} (2012) 143
  [arXiv:1205.5736 [hep-ex]];

\bibitem{mssmhiggs}
 S.~Schael {\it et al.}  [ALEPH and DELPHI and L3 and OPAL and LEP Working Group for Higgs Boson Searches Collaborations],
  Eur.\ Phys.\ J.\ C {\bf 47}, 547 (2006)
  [hep-ex/0602042].

\bibitem{Lees:2012xj}
  J.P.~Lees {\it et al.}  [BaBar Collaboration],
  Phys.\ Rev.\ Lett.\  {\bf 109}, 101802 (2012)
  [arXiv:1205.5442 [hep-ex]].

\bibitem{Spira:1995mt}
M.~Spira,
arXiv:hep-ph/9510347.

\bibitem{LHCHiggs}
https://twiki.cern.ch/twiki/bin/view/LHCPhysics/CrossSectionsFigures\#Higgs\_production\_cross\_sections

%
\bibitem{Harlander:2003ai}
  R.V.~Harlander and W.B.~Kilgore,
  Phys.\ Rev.\ D {\bf 68}, 013001 (2003)
  [hep-ph/0304035].

\bibitem{Chiang:2013ixa} 
  C.~-W.~Chiang and K.~Yagyu,
  arXiv:1303.0168 [hep-ph].

\bibitem{Arbey:2012bp} 
  A.~Arbey, M.~Battaglia, A.~Djouadi, F.~Mahmoudi,
  Phys.\ Lett.\ B {\bf 720}, 153 (2013)
  [arXiv:1211.4004 [hep-ph]].

\bibitem{Gorczyca:2011he} 
  B.~Gorczyca, M.~Krawczyk and ,
  arXiv:1112.5086 [hep-ph].

\bibitem{ATLASnotes}
The ATLAS collaboration, ATLAS-CONF-2013-012, ATLAS-CONF-2013-013, ATLAS-CONF-2013-030, ATLAS-CONF-2013-034.

\bibitem{CMSnotes}
The CMS collaboration, CMS notes CMS-PAS-HIG-13-001, CMS-PAS-HIG-13-002, CMS-PAS-HIG-13-003 and CMS-PAS-HIG-13-003.

\end{thebibliography}
\end{document}